\documentclass[aps, prl, preprint, superscriptaddress, floatfix, amsmath, amssymb]{revtex4}

\usepackage{graphicx}
\usepackage{dcolumn}
\usepackage{bm}
\usepackage{multirow}
\usepackage{subfigure}

\begin{document}

\title{Quantum oscillations of nitrogen atoms in uranium nitride}

\author{A.A. Aczel}
\altaffiliation{author to whom correspondences should be addressed: E-mail:[aczelaa@ornl.gov]}
\affiliation{Quantum Condensed Matter Division, Neutron Sciences Directorate, Oak Ridge National Laboratory, Oak Ridge, TN 37831, USA}
\author{G.E. Granroth}
\affiliation{Quantum Condensed Matter Division, Neutron Sciences Directorate, Oak Ridge National Laboratory, Oak Ridge, TN 37831, USA}
\author{G.J. MacDougall}
\affiliation{Quantum Condensed Matter Division, Neutron Sciences Directorate, Oak Ridge National Laboratory, Oak Ridge, TN 37831, USA}
\author{W.J.L. Buyers}
\affiliation{Chalk River Laboratories, Canadian Neutron Beam Center, National Research Council, Chalk River, Ontario, Canada, K0J 1P0}
\author{D.L. Abernathy}
\affiliation{Quantum Condensed Matter Division, Neutron Sciences Directorate, Oak Ridge National Laboratory, Oak Ridge, TN 37831, USA}
\author{G.D.~Samolyuk}
\affiliation{Materials Science and Technology Division, Physical Sciences Directorate, Oak Ridge National Laboratory, Oak Ridge, TN 37831, USA}
\author{G.M.~Stocks}
\affiliation{Materials Science and Technology Division, Physical Sciences Directorate, Oak Ridge National Laboratory, Oak Ridge, TN 37831, USA}
\author{S.E. Nagler}
\altaffiliation{author to whom correspondences should be addressed: E-mail:[naglerse@ornl.gov]}
\affiliation{Quantum Condensed Matter Division, Neutron Sciences Directorate, Oak Ridge National Laboratory, Oak Ridge, TN 37831, USA}
\affiliation{CIRE, University of Tennessee, Knoxville, TN 37996, USA}

\date{\today}

\begin{abstract}
The vibrational excitations of crystalline solids corresponding to acoustic or optic one phonon modes appear as sharp features in measurements such as neutron spectroscopy. In contrast, many-phonon excitations generally produce a complicated, weak, and featureless response. Here we present time-of-flight neutron scattering measurements for the binary solid uranium nitride (UN), showing well-defined, equally-spaced, high energy vibrational modes in addition to the usual phonons. The spectrum is that of a single atom, isotropic quantum harmonic oscillator and characterizes independent motions of light nitrogen atoms, each found in an octahedral cage of heavy uranium atoms. This is an unexpected and beautiful experimental realization of one of the fundamental, exactly-solvable problems in quantum mechanics.  There are also practical implications, as the oscillator modes must be accounted for in the design of generation IV nuclear reactors that plan to use UN as a fuel. 
\end{abstract}

\maketitle

The spectrum of elementary excitations in materials is one of the core concepts in modern condensed matter physics. The archetypical example is the set of quantized lattice vibrations in crystalline solids, or phonons\cite{ashcroft}. For crystals with more than one atom per unit cell, one expects both acoustic and optic phonon modes and if these are known one can calculate the lattice contribution to fundamental properties such as the heat capacity. The vibrational spectrum at energies above those of the highest optic phonon mode is generally a complicated many phonon continuum that is often weak and featureless. In sharp contrast, for the binary solid uranium nitride (UN)\cite{86_jackman, 84_holden, 05_solontsov, 07_samsel}, where the nitrogen atoms are very light compared to the uranium atoms, our inelastic neutron scattering measurements reveal that the high energy spectrum is greatly simplified and consists of a set of equally-spaced, well-defined modes that can be measured to energies at least ten times as large as that of the optic phonon modes. This data is best explained by assuming that each nitrogen atom behaves as an independent quantum harmonic oscillator (QHO), and the nitrogen motion is therefore a rare experimental realization of an exactly soluble, three-dimensional model in quantum mechanics\cite{book_shankar}. 

\begin{figure}
\centering
\scalebox{0.2}{\includegraphics[angle=0]{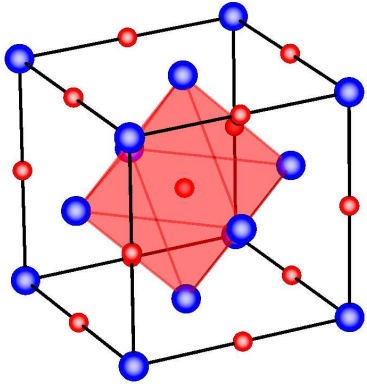}}
\caption{\label{structure} {\bf Rocksalt crystal structure of uranium nitride (UN).} Each N atom (small red spheres) is centered in a regular octahedron of U atoms (large blue spheres).}
\end{figure}

The binary uranium systems of the form UX (X = C, N, S, Se, Te, As, Sb) have relatively simple rocksalt crystal structures as illustrated in fig.~\ref{structure}.  They have been extensively studied due to their vast array of puzzling physical and magnetic properties\cite{84_book}, including unusually high electronic specific heats and drastic suppression of ordered magnetic moments.  Among these systems, UN has received significant attention\cite{10_lu, 11_modak, 11_kotliar, 12_gryaznov, 11_dubovsky} recently due to its potential use as a high-temperature nuclear fuel\cite{09_burkes, 12_schriener, 09_petti}.

\begin{figure*}
\centering
\scalebox{0.18}{\includegraphics{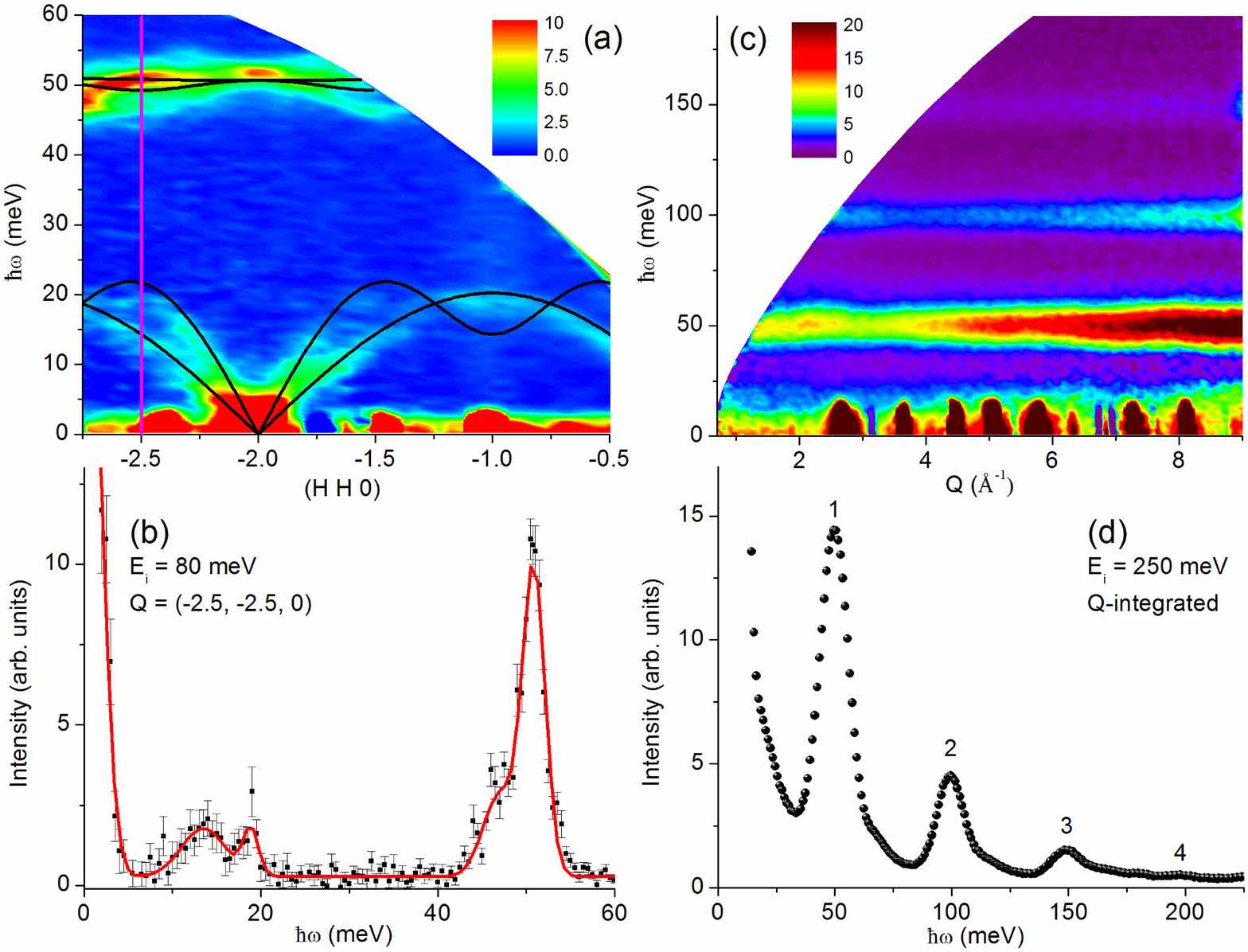}}
\caption{\label{phonons} {\bf Inelastic neutron  scattering data from UN measured at SEQUOIA for T~$=$~5~K.} (a) Color intensity plot of the scattering with $E_i$~$=$~80~meV. The vertical axis $\hbar \omega$ denotes the energy transfer, and the horizontal axis the wavevector transfer along high-symmetry direction ($HH0$). The integration ranges for the perpendicular directions ($H \bar{H} 0$) and ($00L$) were 0.2~reciprocal lattice units (rlu). The data clearly shows the acoustic and optic one phonon modes, and the black curves show fits of the phonon dispersion to the two force-constant model described elsewhere\cite{74_wedgwood}. (b) A constant $Q$-cut taken along the vertical line in (a) with an integration range of 0.2~rlu in the ($HH0$), ($H \bar{H} 0$) and ($00L$) directions. The solid line represents a fit to a superposition of Gaussian functions. (c) Color intensity plot of the $E_i$~$=$~250~meV data summed over all directions in reciprocal space, with the magnitude of the wavevector, $Q$, along the horizontal axis. Several modes are evident, evenly-spaced in energy by approximately 50~meV intervals.  As discussed in the text, these are attributed to quantum harmonic oscillator behaviour of the nitrogen atoms in UN. (d) $Q$-integrated data from (c), clearly showing the oscillator excitations. }
\end{figure*}

The primary magnetic and lattice excitations of UN have been investigated previously via inelastic neutron scattering\cite{86_jackman, 84_holden, 74_wedgwood}.  Despite these efforts, several intriguing open questions remain concerning the details of the experimental spectra.  Since the initial measurements, significant advances have been made in inelastic neutron scattering using time-of-flight methods.  Next generation chopper spectrometers allow for measurements over much broader energy ($\hbar \omega$) and momentum ($\vec{Q}$) transfer ranges than were previously accessible, with both improved intensity and resolution.  For these reasons, the excitations in UN were re-examined using the Fermi chopper spectrometers SEQUOIA\cite{06_granroth, 10_granroth} and ARCS\cite{12_abernathy} at the Spallation Neutron Source of Oak Ridge National Laboratory.  In addition to the expected magnon and phonon modes, this investigation resulted in the unexpected discovery of a series of excitations spaced equally in energy by intervals of approximately 50 meV, and extending up to at least 500 meV.  

\begin{figure*}
\centering
\scalebox{0.18}{\includegraphics{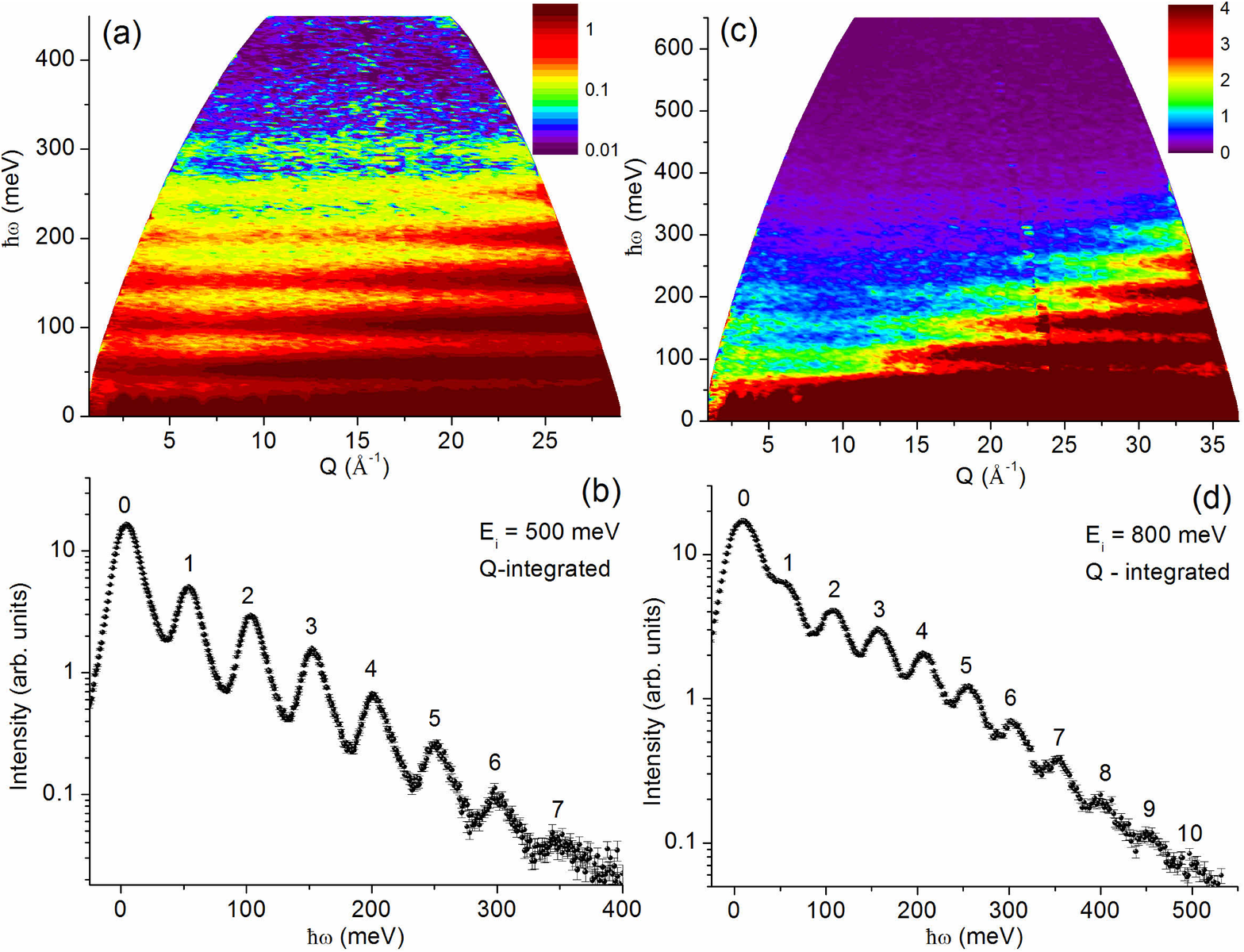}}
\caption{\label{high_energy_data}{\bf Inelastic neutron  scattering data from UN measured at ARCS for T~$=$~5~K.} (a) Color intensity plot of the scattering with $E_i$~$=$~500~meV averaged over all measured crystal orientations.  (b) The scattering in (a) integrated over all $Q$. (c) and (d) are similar plots for $E_i$~$=$~800~meV.  The data shows a clear series of 50~meV evenly-spaced excitations as expected for a quantum harmonic oscillator.  In particular, ten different oscillator excitations ranging in energy from 50 to 500~meV can be seen in (d).}
\end{figure*}

\section{Results}

To perform the measurements, the same UN crystal utilized by Jackman {\it et al}\cite{86_jackman} was aligned with the ($HHL$) plane horizontal and cooled to a temperature T~$=$~5~K.  Figure~\ref{phonons}(a) shows representative results obtained at SEQUOIA using incident neutron energy $E_i$~$=$~80~meV for a typical symmetry direction in reciprocal space.  A constant $Q$-cut through the data, at the position corresponding to the vertical line, is illustrated in fig.~\ref{phonons}(b). The solid red curve is a fit to a superposition of Gaussian peaks at the mode positions. The measured phonon dispersion relations are consistent with earlier investigations\cite{86_jackman, 77_dolling}. The black lines in fig.~\ref{phonons}(a) are obtained from a fit of the data to a simplified model\cite{74_wedgwood} that includes only nearest-neighbour U-N (65~$\pm$~2~N/m) and U-U (50~$\pm$~4~N/m) force constants. This model suffices for illustrative purposes, although more detailed models\cite{86_jackman, 92_jha, 94_jha, 10_lu, 11_modak, 11_kotliar} give a closer match to the observed dispersion. Some of those models also include N-N force constants, although they are found to be negligibly small compared to their U-N and U-U counterparts. 

Notably, the optic modes are well-separated in energy from the acoustic modes and are weakly-dispersive centered around $\hbar \omega$~$\sim$~50~meV\cite{77_dolling, 86_jackman}. Due to the large mass ratio $M/m$, where $M$ and $m$ are the masses of the U and N atoms respectively, the eigenvectors of the modes are such that the U atom motions are largely reflected by the acoustic phonons and the optic modes correspond primarily to motions of the N atoms\cite{74_wedgwood}. 

The upper and lower panels in fig.~\ref{phonons}(c) and (d) show SEQUOIA data taken with $E_i$~$=$~250~meV covering a larger range of energy transfer.  The upper panel shows the data averaged over all measured crystal orientations and the lower panel is a plot of intensity vs. energy transfer for this data summed over $Q$.  The data is striking and reveals a series of essentially dispersionless excitations extending up in energy from the optic mode, and evenly spaced by approximately 50~meV intervals. The intensity of each mode increases with increasing momentum transfer over several~\AA$^{-1}$~indicating that these excitations are vibrational as opposed to magnetic.  These data were extended greatly in both energy and wavevector range by further measurements using the ARCS chopper spectrometer. Figures~\ref{high_energy_data}(a) and (b) show ARCS data with $E_i$~$=$~500~meV plotted on a logarithmic intensity scale. Figures~\ref{high_energy_data}(c) and (d) show data from the same instrument with $E_i$~$=$~800~meV; the upper panel (c) has intensity on a linear scale while (d) is also on a logarithmic scale.  A series of evenly-spaced excitations are clearly visible up to the 10th order at an energy of 500 meV. The well-defined peaks in the orientationally-averaged data imply that the modes are localized and isotropic.  

\begin{figure*}
\centering
\scalebox{0.18}{\includegraphics{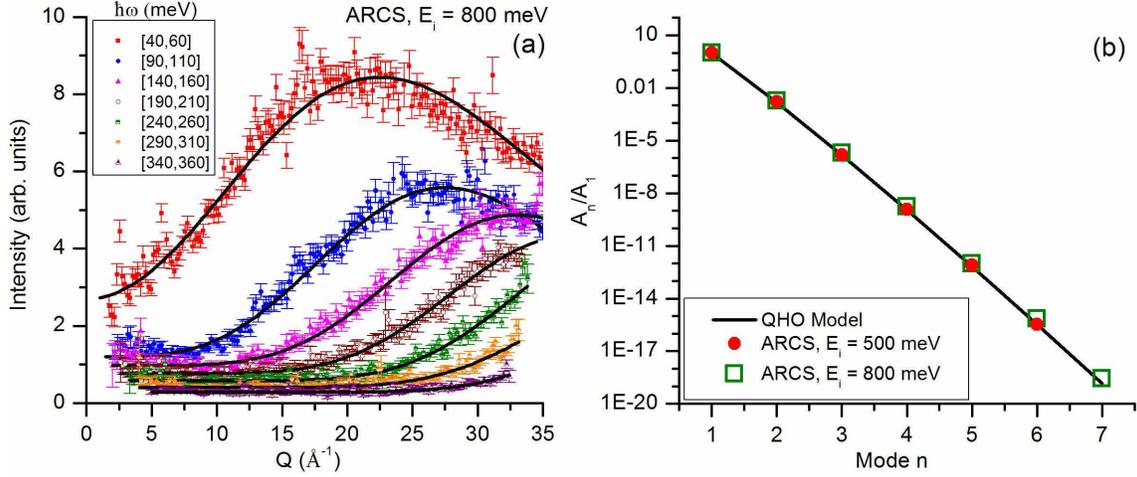}}
\caption{\label{high_energy_fits}{\bf Fitting results and analysis of the UN oscillator modes.} (a) The $Q$-dependence of the intensity for the $n$~$=$~1-7 oscillator modes with $E_i$=800 meV.  The legend indicates ranges of integration over energy.  The solid lines are fits to the QHO model as described in the text. (b) The $A_n$ coefficients for the $n$~$=$~1-7 modes normalized by the values of A$_1$.  The coefficients are extracted from fits to the data as described in the text.  The solid line is the prediction of the QHO model with $\hbar \omega_o = 50$~meV and $m$ corresponding to the mass of a nitrogen atom.  The ratio is plotted on a logarithmic scale and spans almost 20 orders of magnitude.}
\end{figure*}

Some enhanced response at an energy corresponding to twice that of the optic phonon was observed in previous neutron scattering studies of UN and attributed to two-phonon processes\cite{74_wedgwood}.  However, the observation of such an extended series of equally-spaced and well-defined modes is unprecedented. The most obvious system exhibiting a spectrum consisting of equally-spaced modes is the quantum harmonic oscillator (QHO). Here one can postulate that the extremely light mass of the N atom relative to the U atom leads to a situation where high frequency vibrations in the crystal essentially consist of motions of individual N atoms in a very isotropic and harmonic potential arising from a surrounding regular octahedron of U atoms. Each N atom is then a nearly ideal realization of the QHO in three dimensions, with a fundamental frequency corresponding to an energy $\hbar \omega_o$~$\approx$~50 meV, equivalent to a temperature of 580~K. The QHO is one of the simplest and best understood theoretical models in quantum mechanics\cite{book_shankar}, and serves as the foundation for understanding diverse phenomena such as the vibrational modes of molecules, the motion of atoms in a lattice, and the theory of heat capacity. We show below that the QHO model can quantitatively describe the high-energy vibrational features observed in the UN data.

The neutron scattering signal expected from the QHO is well-known\cite{textbook_lovesey}. The dynamical response for $k_{B}$T~$<<$~$\hbar\omega_o$ can be written as:
\begin{equation}
S(Q,\omega) = \sum_{n}S_{n}(Q, \omega)
\end{equation}
where the contribution to the intensity arising from the $n$th oscillator excitation is given by:
\begin{equation}
S_n(Q, \omega) = \frac{1}{n!} \left(\frac{\hbar Q^{2}}{2m\omega_o}\right)^n exp\left(\frac{-\hbar Q^2}{2m\omega_o}\right) \delta(\hbar\omega-n\hbar\omega_o) 
\end{equation}
The argument of the exponent resembles a Debye-Waller factor and reflects the zero-point motion of the oscillator.  It is interesting to note that the intensity maximum for each mode occurs at the $Q$-position corresponding to the recoil scattering for a free particle of mass $m$ (see Supplementary Information).

Equation (2) reflects the intensity expected for a single inelastic scattering event.  Since the modes are evenly spaced, in a real experiment multiple scattering events in any combination of elastic and inelastic scattering will also result in intensity at the positions of the modes. Hence any comparison of the QHO theory to the data must account for the intensity arising from multiple scattering in some fashion. As a first approximation, the measured data for UN was compared to the QHO model by fitting constant energy cuts (20~meV integration range) corresponding to the $n$~$=$~1-7 modes for both $E_i$~$=$~500~meV and 800~meV data to the following functional form:
 \begin{equation}
S_n(Q) = A_nQ^{2n}exp(-CQ^2)+B_n
\end{equation}
Here the multiple scattering contribution to the observed intensity at each mode position is assumed to be a constant $B_n$ independent of momentum $Q$: this is the simplest possible assumption.  

Figure~\ref{high_energy_fits}(a) shows the $Q$-dependence of the seven lowest oscillator modes for the $E_i$~$=$~800~meV data with solid lines indicating the corresponding fits to equation (3).  The fits are excellent over the entire range of measured data and this is also true for the $E_i$~$=$~500~meV data (not shown). Figure~\ref{high_energy_fits}(b) plots the ratios of the fitted parameters $A_n/A_1$ for both data sets. The solid curve indicates the prediction of the QHO model for a nitrogen atom with $\hbar \omega_0$~$=$~50~meV, with the non-integer $n$ values interpolated by using $\Gamma(n+1)$ to calculate $n!$. The values obtained from the data agree very well with the model over nearly 20 orders of magnitude.  The maximum relative deviation for any one ratio is within a factor of two for the $E_i$~$=$~800~meV data and within 20~\% for the $E_i$~$=$~500~meV data.  Given the large range of values covered and the simple assumption made for the multiple scattering, this is strong confirmation that the QHO picture applies to the nitrogen atoms in the uranium lattice. 

The magnitude of the fitted parameters $B_n$ are found to be monotonically-decreasing with energy and dependent on the incident neutron energy. In addition to the intrinsic multiple scattering contribution to the intensity at the mode position the value of $B_n$ may be affected by background terms. The background can arise from several sources, including overlap of the oscillator modes due to imperfect instrumental resolution, and contributions from acoustic-acoustic and acoustic-optic multiphonon excitations. The observation that the high $n$ modes are visible at low $Q$ implies that the intrinsic multiple scattering contribution to $B_n$ is larger than the other background terms. More discussion of the $B_n$ parameter is provided in the Supplementary Information. 

The parameter $C$ is related to the zero point energy of the oscillator.  For a nitrogen atom in a harmonic potential corresponding to $\hbar \omega_o = 50$~meV the calculated value of $C$ is 3.0x10$^{-3}$~\AA$^{2}$.  The mean value of $C$ determined from all modes at both incident energies was found to be 2.75(10)x10$^{-3}$~\AA$^{2}$; this is very close to the calculated value. As a further check, the data for each $E_{i}$ was analyzed using a global fit (see Supplementary Information) obtaining values for $C$ of 2.76(2)x10$^{-3}$~\AA$^2$ and 2.64(2)x10$^{-3}$~\AA$^2$ for the $E_i$~$=$~500~meV and 800~meV data sets respectively, also in good agreement with expectations. The slight discrepancies between the fitted and calculated $C$ values may be due to the simple assumption made for the multiple scattering.

Clearly, the QHO model gives an outstanding description of the data. It should be noted that a calculation of the multiphonon scattering for a single frequency, perfectly-dispersionless Einstein model leads to a result that is equivalent to the QHO model\cite{book_bunkov}. However, the oscillator picture is a simpler and more physically reasonable description for the localized vibrations of N atoms in UN, particularly when one considers that it is not an ideal Einstein solid: the crystalline environment and acoustic modes also affect the scattering. Detailed theoretical predictions exist for the response of a binary harmonic solid consisting of a lattice of diatomic molecules with a light atom of mass $m$ bound to a heavier atom of mass $M$\cite{83_lovesey}. In the limit $m << M$, appropriate for UN, the finite mass of the U atoms can be accounted for approximately by replacing the delta function in equation~(2) with a Gaussian:
\begin{equation}
exp\left(\frac{-(\hbar\omega-n\hbar\omega_o-\frac{\hbar^2 Q^2}{2M})^2}{2\sigma^2}\right)
\end{equation}
This modification suggests that the oscillator modes should be broadened to a width $\sigma$ and the position of the $n$th mode shifted by the uranium recoil to $\omega$~$=$~$n\omega_o+\frac{\hbar^2 Q^2}{2M}$.  For UN, it is not possible to associate a specific N atom with an individual U atom, nevertheless it is still reasonable to compare the data with the binary solid model. To assess how well this model describes the present data, constant-$Q$ cuts (2~\AA$^{-1}$~integration range) from the ARCS $E_i$~$=$~500~meV measurement were fit to a superposition of several Gaussian functions. The net result of the fitting shows that the modes have an average intrinsic full-width half-maximum of 26 meV with no systematic behavior evident as a function of $Q$ or $n$ (see Supplementary Information). The magnitude of the broadening is in agreement with the prediction of the binary solid model\cite{83_lovesey} that the intrinsic energy width of the $n > 1$ peaks should be of the same order as the bandwidth of the acoustic phonon modes. Inspection of figure 2(a) shows this width to be slightly over 20~meV. Except for a possible small effect in the $n=1$ mode, the fitting does not reveal a measurable shift of the mode energies as a function of $Q$.  Multiple scattering contributions at the mode positions may mask any shift at higher $Q$. 

\begin{figure}
\centering
\scalebox{0.11}{\includegraphics[angle=0]{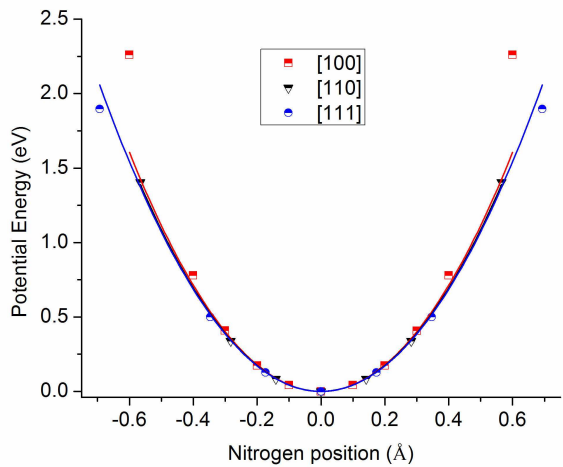}}
\caption{\label{DFT} {\bf DFT calculated potential energy of the N atoms in the system UN.} The potential energy is shown for atomic displacements relative to the equilibrium position for each of the three indicated symmetry directions as indicated by the symbols. The solid lines are fits of the calculated small displacement limit values to parabolic (i.e. harmonic) potentials.  The potential energy is very isotropic and harmonic over a wide range.  Deviations are visible above 1 eV, especially along the [100] direction.}
\end{figure}

In order to investigate why the isotropic QHO description is so appropriate for UN, density functional theory (DFT) has been used to calculate the potential energy of the nitrogen atoms relative to their equilibrium positions for displacements along major crystallographic directions. Although DFT often fails to give the true ground state for correlated materials, in fact it has been shown to give a good description of the electronic energies in UN\cite{11_kotliar}. Moreover, DFT gives highly reliable values for relative energy differences of displaced atoms independent of the precise ground state. The DFT results shown in fig.~\ref{DFT} verify that the potential is harmonic over a wide energy range, certainly for $E$~$<$~1~eV.  At the same time, in this range the potential energy along different directions is isotropic to within 2~\%. The ab-initio calculations predict an energy spacing for the oscillator modes of 50~meV, corresponding very well to the spacing of the excitations observed in the neutron scattering measurements. 

\section{Discussion}

The QHO behavior observed here should be visible in many materials where the constituent atoms have greatly different masses. It is also important that there be weak interactions between light atoms sitting in a very harmonic potential.  Possible harmonic oscillator behavior has long been sought and investigated in metallic hydrides, especially for zirconium and titanium systems\cite{57_andresen, 59_whittemore, 86_kolesnikov, 92_kolesnikov, 94_kolesnikov, 86_ikeda}. However, the potentials at the hydrogen sites generally show significant anisotropy or anharmonic effects\cite{71_couch, 97_elsasser}. These features manifest themselves as fine structure and uneven spacing in the hydrogen vibrational modes, and they arise from H-H interactions, crystalline anisotropy, and low potential barriers to hopping of the protons. In sharp contrast, the potential at the nitrogen sites of UN is very harmonic and isotropic, leading to evenly-spaced nitrogen vibrational modes exhibiting no fine structure within the instrumental energy resolution. To the best of our knowledge, each nitrogen atom in UN is a unique manifestation of a single atom QHO.

The scattering of neutrons from the oscillator modes may be an important factor in the consideration of UN as a nuclear fuel material.  Uranium nitride has several desirable characteristics for this application, including excellent thermal conductivity and a high melting temperature.  For that reason, it is currently under intense consideration for use in proposed Generation IV nuclear reactors\cite{04_butler, 09_burkes, 09_petti, 12_schriener}, which operate at high temperatures (500-1000$^\circ$C), allowing for improved efficiency in the conversion of heat to electricity.

Nuclear reactor designs require a detailed knowledge of the neutron cross-sections and thermal behavior of the constituent atoms composing the nuclear fuel. Fermi\cite{36_fermi} showed that harmonic oscillations of protons in metals have a significant measurable effect on total neutron cross-sections. In addition to affecting the total cross-section, the oscillator modes in UN could also impact operation via self-moderation of neutrons.  The modes will be thermally-populated at the proposed Generation IV operating temperatures and it is necessary to understand if there will be any impact on the thermal conductivity. Given the strong scattering processes shown in the present work, it is clear that the oscillator excitations must be accounted for properly in any detailed Generation IV designs. 

\section{Methods}
For the neutron scattering experiments, the UN single crystal was sealed in an ultrathin Al sample can\cite{11_stone} with He exchange gas and cooled to approximately 5~K in a closed cycle He refrigerator. For the SEQUOIA\cite{10_granroth}  measurements, a Fermi chopper with a slit spacing of 3.6~mm and a radius of curvature of 1.53~m was spun at 300~Hz and 360~Hz for $E_i$~$=$~80~meV and 250~meV respectively, yielding FWHM energy resolutions of 4 and 6~\% of $E_i$ at the elastic line. The data was accumulated by combining measurements taken with several different incident neutron directions in the $[HHL]$ plane.  The data from ARCS was obtained using a chopper with an identical radius of curvature and outer dimensions to the SEQUOIA Fermi chopper described above, but with a slit spacing of 0.5~mm. The chopper was spun at 480~Hz ($E_i$= 500 meV) or 600~Hz for ($E_i$= 800~meV) to yield a FWHM energy resolution of 3.5~\% of $E_i$ at the elastic line. The sample was rotated continuously between the $[HH0]$ and $[00L]$ directions during data collection. In all measurements, the data was normalized to account for variations of the detector response and solid angle coverage with a vanadium standard.  This procedure enabled an accurate determination of the orientationally-averaged scattering response.

The DFT calculations used the Quantum Espresso package\cite{09_giannozzi}. Additional technical details of these calculations are presented in the Supplementary Information. 

\section{Acknowledgments}
The authors acknowledge A.I. Kolesnikov, J. Carpenter, E. Iverson, and R.J. McQueeney for useful discussions and A.T.~Savici, T.E.~Sherline and M.J. Loguillo for technical support. This research was supported by the US Department of Energy, Office of Basic Energy Sciences.  A.A.A., G.E.G., G.J.M., D.L.A., and S.E.N. were supported by the Scientific User Facilities Division. G.D.S. and G.M.S. were supported by the Center for Defect Physics and Energy Frontier Research Center. Experiments were performed at the Spallation Neutron Source which is sponsored by the Scientific User Facilities Division.

\section{Author Contributions}
A.A.A., G.E.G., G.J.M., W.J.L.B., D.L.A., and S.E.N. participated in the neutron scattering experiments. The density functional theory calculations were performed by G.D.S. and G.M.S. The manuscript was written by A.A.A. with substantial input from G.E.G. and S.E.N. Critical manuscript comments were provided by all authors. 

\section{Additional Information}
The authors declare no competing financial interests.

\clearpage

\section{\label{sec:level1}{Supplementary Information}}

\section{\label{sec:level1}{Additional Analysis Details of the Oscillator Modes}}

As discussed in the manuscript, constant $E$-cuts of the ARCS $E_i$~$=$~500~meV and 800~meV data corresponding to the $n$th oscillator mode were fit to the functional form: 
\begin{equation}
S_n(Q) = A_nQ^{2n}exp(-CQ^2)+B_n
\end{equation}
These fits placed no constraints on $A_n$ and $C$, and so these parameters were allowed to vary with $n$ and independently of one another. To compare the fitting results to the quantum harmonic oscillator (QHO) model, it was necessary to normalize $A_n$ with respect to $A_1$  to account for a common scale factor in the neutron data. The normalized $A_n$ values agree well with the QHO model\cite{textbook_lovesey} and all of the $C$ values were determined to be within 20~\% of one another, with the average value obtained from fits to both the $E_i$~$=$~500~meV and 800~meV data given by 2.75(10)x10$^{-3}$~\AA$^2$. 

\begin{figure}
\centering
\scalebox{0.19}{\includegraphics{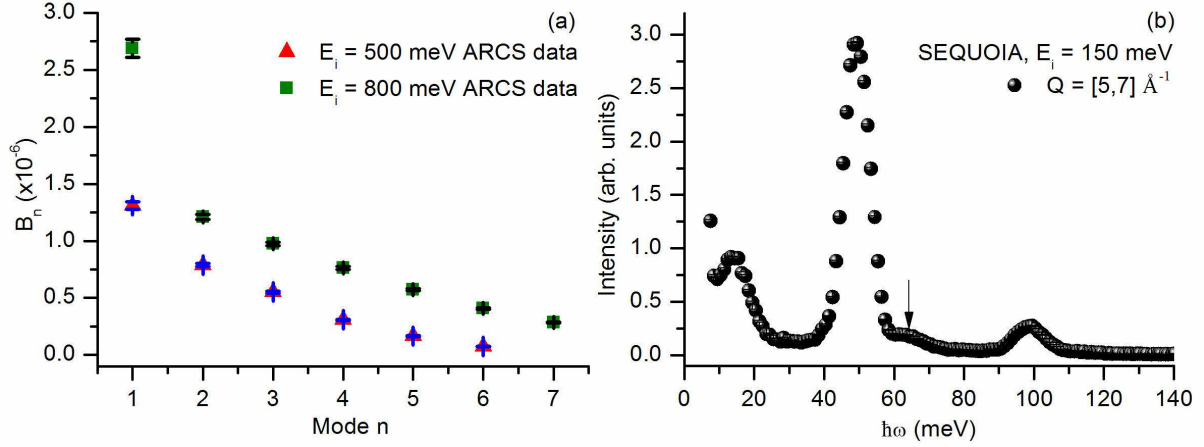}}
\caption{\label{S1}(a) The $B_n$ parameters as determined from fitting the ARCS $E_i$~$=$~500 and 800 meV data to the functional form given by equation (3) in the manuscript. (b) A constant $Q$-cut of $E_i$~$=$~150~meV data from SEQUOIA reveals a weak mode around 65~meV indicated by the arrow, corresponding to a multiphonon excitation from an acoustic zone boundary plus an optic phonon mode. This demonstrates that multiphonon excitations involving the acoustic modes are contributing to the background.}
\end{figure}

Although including the $B_n$ term was necessary for fitting the data, it is not explained by the QHO model. The magnitude of this term is plotted in fig.~\ref{S1}(a) for ARCS data with incident energies of both $E_i$~$=$~500 and 800~meV, and it accounts for both a background and an intrinsic contribution to the oscillator excitation. The background term arises from overlap of the oscillator modes due to imperfect instrumental resolution, plus a contribution from multiphonon excitations involving only the acoustic modes or some combination of the acoustic and optic modes.  The resolution overlap is illustrated by two features in the data.  First, the decreasing $B_n$ values with increasing $n$ arise from the fact that the resolution is a monotonically-decreasing function of energy transfer\cite{12_abernathy}. Second, the energy resolution is also finer for smaller $E_i$, thus explaining the decrease of the $B_n$ parameters with decreasing $E_i$. The multiphonon contribution is illustrated most clearly by considering SEQUOIA data with $E_i$~$=$~150~meV averaged over all measured crystallographic orientations and plotted in fig.~\ref{S1}(b). A series of peaks are visible at 15~meV, 50~meV, 65~meV, and 100~meV. The lowest peak corresponds to the acoustic zone boundary phonon modes, while the peaks at 50~meV and 100~meV correspond to the $n$~$=$~1 and 2 oscillator modes. The weak peak at 65~meV is a multiphonon excitation corresponding to a zone boundary acoustic plus optic phonon mode, but cannot be clearly observed in spectra with higher incident energies due to coarser instrumental energy resolution. Most importantly, multiple scattering due to any combination of elastic scattering and inelastic scattering events from oscillator modes results in intensity at higher order oscillator mode positions.  This plausibly leads to a $Q$-independent contribution to intensity at the position of the $n$th excitation.   The experiments used a rather large single crystal sample originally designed for weak magnetic scattering experiments\cite{84_holden}. Simple considerations taking into account the size of the crystal and the total scattering cross-sections for U$^{238}$ and N confirm that significant multiple scattering is present.  The $Q$ dependence of the multiple scattering should be much weaker and governed by a relatively constant functional form, one much different than the QHO. Next order dependencies of the multiple scattering are difficult to model and are irrelevant to the conclusions of this work. For these reasons, treating this contribution as $Q$-independent serves as a reasonable approximation.  

\begin{figure}
\centering
\scalebox{0.18}{\includegraphics{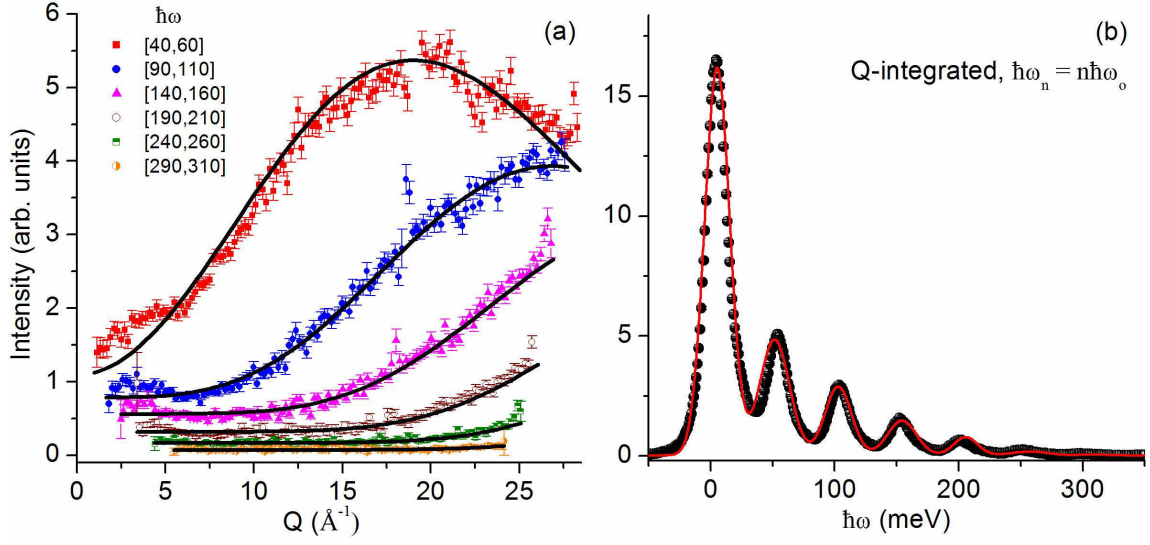}}
\caption{\label{S2} (a) The $Q$-dependence of the $n$~$=$~1-6 oscillator modes and the corresponding fits to the quantum harmonic oscillator model for the $E_i$~$=$~500~meV ARCS data, using the alternative global fitting method described in the text. (b) The same data set integrated over all $Q$ and fit to a superposition of Gaussian functions with the energy of the $n$th mode constrained to be $n \hbar \omega_o$. }
\end{figure}

Another way of fitting the constant $E$-cuts is to assume that $A_n$ and $C$ are directly related to one another as described by the QHO model. Specifically, $A_n$~$=$~$\frac{C^n}{n!}$ and $C$~$=$~$\frac{\hbar}{2m\omega_o}$, where $m$ is the nitrogen mass and $\hbar \omega_o$ is the fundamental energy of the QHO. One can then fit all of the oscillator modes with two common parameters, namely $C$ and a scale factor, along with a different $B_n$ parameter for each mode. Using this alternative method, the $C$ parameter was found to be 2.76(2)x10$^{-3}$~\AA$^2$ and 2.64(2)x10$^{-3}$~\AA$^2$ for the $E_i$~$=$~500~meV and 800~meV data sets respectively. A series of constant $E$-cuts taken from the $E_i$~$=$~500~meV data with the corresponding fits from this second method superimposed are shown in fig.~\ref{S2}(a). The excellent agreement between the two methods demonstrates that the results for the fitting parameters are robust. However, the fitted $C$ values are somewhat lower than the expected value of 3x10$^{-3}$~\AA$^2$ obtained by assuming $\hbar \omega_0$~$=$~50~meV and an atomic mass of 14.  It is possible that some of this discrepancy may be related to the simple assumption made for the $B_n$ terms described above. 

An experimental value of $\hbar \omega_o$ can be determined by directly fitting the mode positions, since these will be largely unaffected by multiple scattering. An overall average estimate can be obtained by fitting $Q$-integrated data to a superposition of Gaussian functions with the energy of the $n$th mode constrained to be $n\hbar\omega_o$; a representative fit of this type to the ARCS $E_i$~$=$~500~meV data is shown in fig.~\ref{S2}(b). This procedure yielded an average value of $\hbar \omega_0$~$=$~50.2(3) meV for the SEQUOIA data and 51.7(1) meV for the ARCS data. The small discrepancy in $\hbar \omega_o$ obtained from the two spectrometers is most likely due to the difference in the integration range of $Q$. The ARCS measurements accessed much higher $Q$ values, and since the position of the $n~=~1$ mode was found to increase with $Q$ as explained below, this should cause a small increase in $\hbar \omega_o$ for the $Q$-integrated ARCS data as compared to the SEQUOIA data.  


\clearpage

\section{\label{sec:level1}Binary Solid Corrections to the QHO Model}

\begin{figure}
\centering
\scalebox{0.13}{\includegraphics{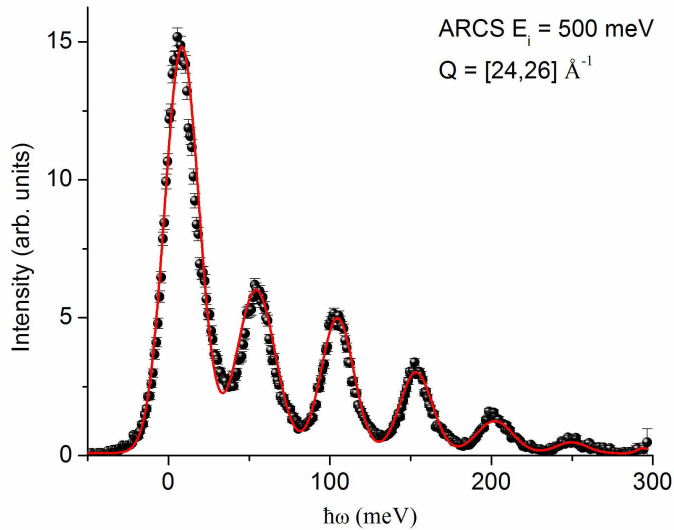}}
\caption{\label{S3} A representative plot of the $E_i$~$=$~500~meV ARCS data with the corresponding fit to a superposition of Gaussian functions. }
\end{figure}

\begin{figure}
\centering
\scalebox{0.18}{\includegraphics{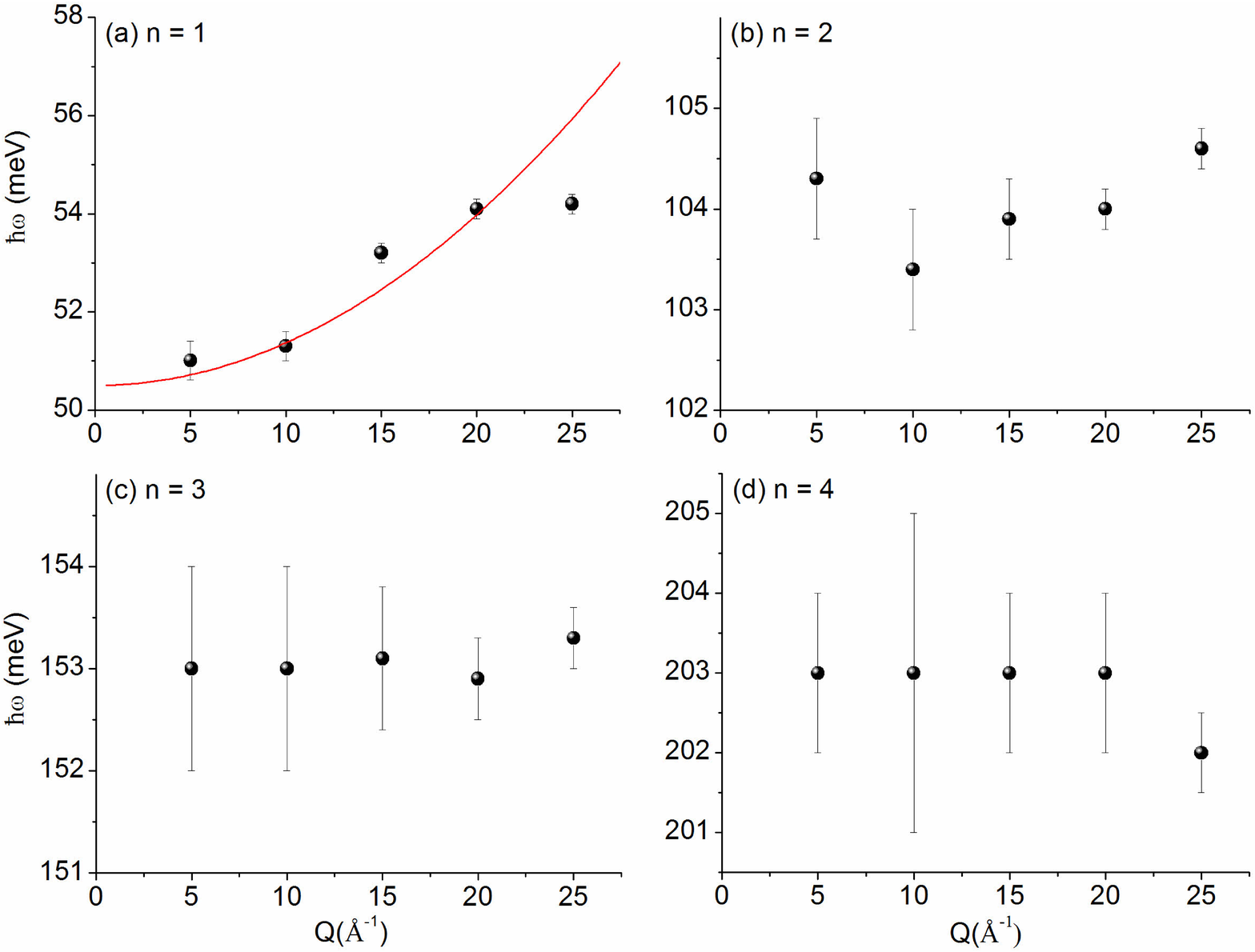}}
\caption{\label{S4}The $Q$-dependence of the positions of the (a) $n$~$=$~1, (b) $n$~$=$~2, (c) $n$~$=$~3, and (d) $n$~$=$~4 modes from the $E_i$~$=$~500~meV ARCS data. The solid curve in (a) corresponds to the position shift of the $n$~$=$~1 mode predicted by the binary solid model. The position of this mode shows an increase with $Q$ roughly proportional to the uranium recoil line as expected, while the $n$~$>$~1 mode positions show no variation within experimental error. }
\end{figure}

As discussed in the manuscript, theory\cite{83_lovesey} predicts that the oscillator modes of a real binary crystal will be broadened and can be approximated by Gaussian functions. In principle, the width of the peaks depends on several factors, including the temperature, wavevector, and the ratio of the fundamental oscillator mode energy to the zone boundary energy of the acoustic phonon modes. The width involves a convolution of the acoustic modes and harmonic oscillator response and therefore should be of the same order as the bandwidth of the acoustic modes.  The model also predicts a shift of the oscillator mode positions by an amount corresponding to the uranium recoil scattering.  These predictions were investigated by fitting constant $Q$-cuts of the data (2~\AA$^{-1}$ integration range) as well as $Q$-integrated cuts to a superposition of Gaussian functions with independently varying positions and widths. A representative plot of the $E_i$~$=$~500~meV ARCS data with the corresponding fit is shown in fig.~\ref{S3}, and some fitting results are presented in Table I and fig.~\ref{S4}.  Table I shows that the spacing of the modes agrees well with the estimates of $\hbar \omega_0$ obtained by the constrained fitting as described above. The oscillator modes were also found to be characterized by intrinsic full-width-half-maxima (FWHM) of $\approx$~26~meV as shown in Table~I (assuming that the resolution and sample contributions to the widths add in quadrature) with no systematic $Q$ or $n$-dependence. For the peaks with $n > 1$ the values are in good agreement with the binary solid model prediction, since the acoustic phonon bandwidth in UN is around 20~meV. On the other hand, while the position of the $n$~$=$~1 mode increases with $Q$ roughly according to the uranium recoil scattering as predicted, the higher order mode positions show no $Q$-variation within the error bars as indicated by both Table~I and fig.~\ref{S4}. Therefore, it was not necessary to account for this shift in the analysis of the constant $E$-cuts described above. This deviation from the prediction of the binary solid model may be due to multiple scattering effects that have already been discussed in detail.

It is worth noting that for the $Q$-integrated data fitting of the width and position of the optic ($n = 1$) mode is systematically affected by its proximity to the large peak near zero energy. The actual width of the $n = 1$ mode should be determined by the bandwidth of the optic mode.  In addition, an apparent shift in the average position of the $n = 1$ mode as a function of $Q$ may be due to a combination of dispersion and lack of full orientational-averaging for small $Q$.

\begin{table*}[h]
\caption{The peak positions of the oscillator modes for selected $Q$-ranges in \AA$^{-1}$ and the fitted and intrinsic $Q$-integrated FWHM in meV from the $E_i$~$=$~500~meV ARCS data. The instrumental energy resolution is also presented for comparison and was calculated according to \cite{12_abernathy}.}
\centering
   \begin{tabular}{| c | c | c | c | c | c | c | c |}
      \hline 
\multirow{2}{*}{\textbf{Mode}} & \multicolumn{3}{c}{\textbf{Mode Position}} \vline & \multicolumn{3}{c}{\textbf{$Q$-integrated FWHM}} \vline\\   \cline{2-7}
& \textbf{$Q$~$=$~[4-6]} & \textbf{$Q$~$=$~[14-16]} & \textbf{$Q$~$=$~[24-26]} & \textbf{Fitted} & \textbf{Resolution} & \textbf{Intrinsic}  \\  \hline
1 & 51.0 (4) & 53.2(2) & 54.2(2) & 30.1(5) & 15 & 26.1(6)   \\  \hline
2 & 104.2(6) & 103.9(4) & 104.5(2) & 25.8(6) & 13 & 22.3(7)    \\  \hline
3 & 153(1) & 153.1(7) & 153.3(3) & 29(1) & 12 & 26(1) \\  \hline
4 & 203(1) & 203(1) & 202.0(5) & 30(2) & 11 & 28(2)    \\  \hline
5 & 252(2) & 253(2) & 250(2) & 29(3) & 9 & 28(3)  \\ \hline
6 & - & 301(3) & - & 27(10) & 8 & 26(10)  \\ \hline
   \end{tabular}
\end{table*}

\clearpage

\section{\label{sec:level1}Large Incident Energy Limit}

\begin{figure}[h]
\centering
\scalebox{0.14}{\includegraphics{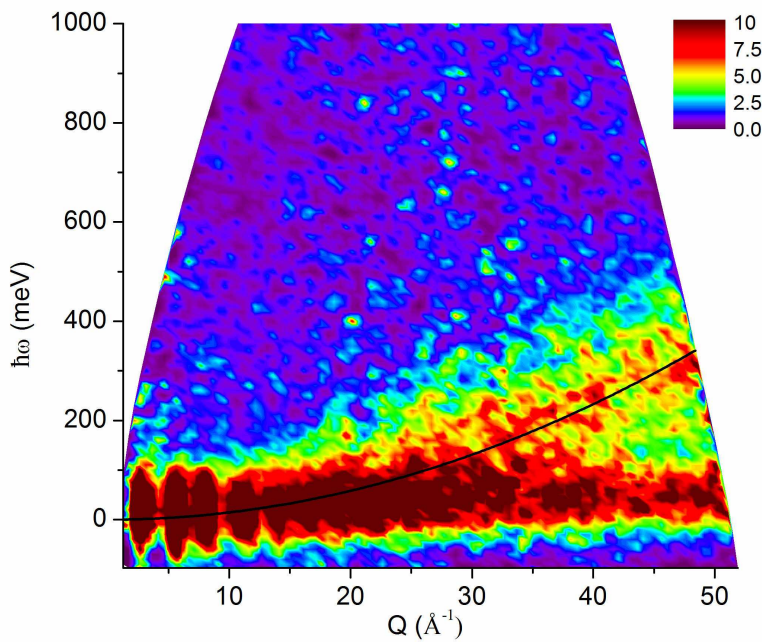}}
\caption{\label{S5} Inelastic neutron scattering spectrum from ARCS with E$_i$~$=$~1.5~eV showing that the response corresponds well to nitrogen recoil scattering in this regime. The black curve is the calculated nitrogen recoil line.}
\end{figure}

The position of maximum intensity of the $n$th oscillator mode in $Q-E$ space corresponds to $E = \frac{\hbar^2 Q^2}{2 m}$, and this is equivalent to the expression for the nitrogen recoil energy. In the limit of large $E_i$, the individual oscillator modes are not resolved due to the coarse energy resolution associated with these incident energies, and the resulting overlapping modes clearly trace out the nitrogen recoil curve. To demonstrate this, ARCS data with $E_i$~$=$~1.5~eV, a Fermi chopper frequency of 600~Hz, and a FWHM energy resolution of 3.5~\%~of~$E_i$~$=$~52.5~meV at the elastic line is shown in fig.~\ref{S5}. The nitrogen recoil is superimposed on the data as indicated by the black curve.  

\clearpage

\section{\label{sec:level1}Density Functional Theory Calculations}

The electronic structure within the density functional theory (DFT) was obtained using the Quantum ESPRESSO package\cite{09_giannozzi}. The calculation was performed using a plane-wave basis set and ultrasoft pseudopotential\cite{90_vanderbilt} in a RKKJ scheme\cite{90_rappe}. The uranium pseudopotential was obtained from an ionized electronic configuration 6$p^6$6$d^1$5$f^3$7$s^1$ with cutoff radii equal to 3.5 atomic units (au), 1.7 au, 2.6 au and 1.6 au for $s$, $p$, $d$, and $f$ angular momentum. The electronic levels deviate from the all-electron ones by less than 0.1~meV. We used the Perdew, Burke and Ernzerhof\cite{96_perdew} exchange-correlation functional. The Brillouin zone (BZ) summations were carried out over a 4 x 4 x 4 BZ k-points grid for a supercell containing 64 atoms. The electronic smearing with a width of 0.02 Ry was applied according to the Methfessel-Paxton method. The plane wave energy and charge density cut-offs were 73 Ry and 1054 Ry respectively, corresponding to a calculation accuracy of 0.2 mRy/atom. The nitrogen atom potential was obtained from the total energy modification of a 2 x 2 x 2 supercell when one nitrogen atom was shifted from the equilibrium position in the [100], [110] or [111] directions and the remaining atoms were held fixed in their equilibrium positions.

\end{document}